\documentclass{article}

\def\abstract#1{\vskip 7mm 
        \begin{center}{\large Abstract}\par \smallskip
                \begin{minipage}[c]{12cm}
                        \small #1
                \end{minipage}
        \end{center}
}
\def\title#1{\begin{center}{\Large\bf #1}\end{center}}
\def\author#1{\vskip 5mm \begin{center}{#1}\end{center}}
\def\address#1{\begin{center}{\it #1}\end{center}}
\makeatletter
\def\vereq#1#2{\lower3pt\vbox{\baselineskip1.5pt \lineskip1.5pt
\ialign{$\m@th#1\hfill##\hfil$\crcr#2\crcr\sim\crcr}}}
\makeatother

\usepackage{graphicx}
\usepackage{amssymb}
\usepackage[mathscr]{eucal}

\newcommand{\gs}{g_{\mathrm{s}}}
\newcommand{\Nflux}{\mathcal{N}}
\newcommand{\phistrg}{\phi_{\mathrm{strg}}}
\newcommand{\phiuv}{\phi_{_{\mathrm{UV}}}}
\newcommand{\alphas}{\alpha'}
\newcommand{\mpl}{m_{\mathrm{Pl}}}
\newcommand{\phiini}{\phi_{\mathrm{in}}}
\newcommand{\phiend}{\phi_{\mathrm{end}}}
\newcommand{\phieps}{\phi_{\epsilon}}
\newcommand{\phiDBI}{\phi_{_{\mathrm{DBI}}}}

\begin{document}

\title{%
  Constraints on brane inflation from WMAP3
}

\author{Larissa Lorenz\footnote{lorenz@iap.fr}}
\address{$^{1}$Institut d'Astrophysique de Paris, UMR
7095-CNRS, Universit\'e Pierre et Marie Curie,\\98bis boulevard Arago,
75014 Paris, France}
  


\abstract{
Considerable and ongoing effort is made to identify promising scalar field candidates in string theory to drive a cosmological period of inflation. At stake is the possibility that fundamental string parameters could be encoded in observables such as the CMB perturbation spectrum. In this contribution, we hold a concrete model of string inflation (KKLMMT) up against WMAP3 and discuss the constraints obtained.
}

\section{Introduction}
In recent years, the hope of embedding cosmological inflation into superstring theory has been put on more solid grounds. While they remain challenging, issues such as moduli stabilization are better understood, and scenarios for both open and closed string mode inflatons have been constructed. With its tight relation to observables of current and future CMB experiments, inflation could provide the decisive missing link between string theory and observation. We investigate if the WMAP3 data provides constraints on the parameters of one particular (open string) scenario, known as the KKLMMT model of brane inflation \cite{kklmmt}. To this end, we identify its cosmological parameters and how they relate to the underlying string geometry, followed by a comparison to the WMAP3 data using numerical integration of the perturbations and MCMC methods \cite{ourpaper} (see also \cite{tye}).

\section{Setting the stage in string theory}
The KKLMMT inflaton field $\phi=\sqrt{T_{3}}r$ corresponds to  the distance $r$ between a D3 and an anti-D3 brane in a 10d supergravity background. $T_{3}$ denotes the brane tension, $T_{3}= 1/[(2\pi)^{3} \gs \alpha'^{2}]$, with string coupling $\gs$ and $\alpha'=l_{\mathrm{s}}^{2}$ the string length squared. To understand the dynamics of $\phi$ and calculate its potential, one has to start from the 10d action of type IIB superstring theory and find solutions for the metric and all $n$-forms. A supergravity metric ansatz reads ${\rm d}s^{2}=h^{-1/2}(r)g_{\mu\nu}
{\rm d}x^{\mu}{\rm d}x^{\nu}+h^{1/2}(r){\rm d}s_{6}^{2}$,
i.e. a 4d
extended space-time (along the worldvolume of the branes) and six compactified dimensions. 
The function $h(r)$ is called the warp factor. For the 6d section, the choice of interest (in view of the desired cosmological outcome) is
${\rm d}s_{6}^{2}={\rm d}r^{2}+r^{2}{\rm d}s_{T_{1,1}}^{2},$
with \({\rm d}s_{T_{1,1}}^{2}\) the conifold metric \cite{Candelas:1989js}. To enforce 
``warping'' on \(T_{1,1}\), non-vanishing
background fluxes (which are characterized by an integer number $\Nflux$ \cite{Klebanov:2000hb}) are given to certain $n$-forms. This geometry is called the Klebanov-Strassler (KS) throat, defined by the throat's bottom $r_{0}$, edge $r_{_{\mathrm{UV}}}$ (where it is glued into the rest of the 6d manifold), and a dimensionless parameter $v$, measuring the relative size of the 5d conifold base. 
The KS throat is a an explicit example\footnote{where notably $v$ is fixed at $v=16/27$}, but one may consider generic ``warped throats'' which appear in many flux compactifactions.

The (heavy) anti-D3 brane is embedded at \(r_{0}\) within this deformed background; its presence adds a small warp factor perturbation $\delta h(r,r_{0})$. The (light) D3 brane probes the resulting 
geometry: Inserted at \(r_{1}\gg r_{0}\) (far from
the bottom, but below the edge $r_{_{\mathrm{UV}}}$), it experiences gravity
and Ramond-Ramond interactions with the anti-D3  through closed string modes. 
The radial inter-brane distance \(r=r_{1}-r_{0}\) 
is interpreted as the inflation field \(\phi\) (up to normalisation), and its potential \(V(\phi)\) is calculated from the Coulomb-like force in the limit $r \gg l_{\mathrm{s}}$ \cite{kklmmt}. Inflation proceeds while the D3 approaches the anti-D3, hence $\phi$ decreases. At a critical $\phistrg$, when the branes' proper distance equals $l_{\mathrm{s}}$, a tachyon (the lightest open string mode) 
appears, and $V(\phi)$ calculated from closed
mode exchange is no longer valid. The two branes then annihilate in a complex process followed by the reheating era. 

The resulting effective four-dimensional action for the inflaton field $\phi$ in this model reads
\begin{equation}\label{action}
S =  -\frac{1}{2\kappa} \int R
\sqrt{-g} \, \mathrm{d}^4 x - \int \left[T(\phi)\,
  \sqrt{1+\frac{1}{T(\phi)} g^{\mu\nu} \partial_\mu \phi
    \partial_\nu \phi } +T(\phi)\right]\sqrt{-g}\, \mathrm{d}^4 x,
\end{equation}
where $\kappa \equiv 8\pi/\mpl^2$,
and $T(\phi)=T_{3}/\tilde{h}(\phi)$ is the position-dependent brane tension. $\tilde{h}(\phi)$ includes the anti-D3's perturbation and follows from the 10d
Einstein equations. $T(\phi)$ represents an upper bound on the field's velocity; while $\dot{\phi}^{2} \ll T(\phi)$, 
one may expand the squareroot to obtain an action with standard kinetic term. 
The true field dynamics, however, are given by the stringy DBI expression in (\ref{action}). A tool to quantify the DBI impact is the ``Lorentz factor'' $\gamma(\phi,\dot{\phi})=[1-\dot{\phi}^{2}/T(\phi)]^{-1/2}$,
with $\gamma\approx 1$ in the standard phase, and $\gamma\gg1$ when the stringy kinetic term is crucial \cite{silverstein}. In the expansion of (\ref{action}), it is easy to  identify the potential (using the explicit form of $\tilde{h}(\phi)$, see \cite{kklmmt}):
\begin{equation}\label{potential}
V(\phi)=2T(\phi)=\frac{M^{4}}{1+\left(\mu/\phi\right)^{4}}\simeq M^{4}\left[1-\left(\frac{\mu}{\phi}\right)^{4}\right]
\end{equation}
The last expression is obtained for $\phi\gg\mu$. This potential is characterized by the overall scale of inflation $M$, and the relative scale $\mu$ for $\phi$. Hence, together with $\phiuv$ (below which the evolution must start) and $\phistrg$ (where brane annihilation sets in), they give a set of four parameters. On the microscopic level, however, $(M,\mu,\phiuv,\phistrg)$ derive from the stringy quantities $(\gs,\alphas,M,v,\Nflux)$\footnote{where $\mu^{4}=\phi_{0}^{4}/\Nflux$, $M^{4}=4\pi^{2}v\phi_{0}^{4}/\Nflux$}.

\section{The standard inflation viewpoint and stringy aspects}\label{stringaspects}
Starting at some initial value $\mu\ll\phiini<\phiuv$, the inflaton moves across a period of standard inflation on the very flat potential (\ref{potential}). The slow-roll approximation can be used until the field reaches $\phieps$; in usual inflation, this means the end of accelerated expansion\footnote{More precisely, we can distinguish $\phi_{\epsilon_{1}},\phi_{\epsilon_{2}}$ (the end of inflation vs. the end of slow-roll), where we find $\phi_{\epsilon_{2}}>\phi_{\epsilon_{1}}$.}. There are, however, new stringy ingredients in the picture of (\ref{action}): The kinetic term of $\phi$ is DBI, and hence Friedmann and Klein-Gordon equations are different from standard (though they reduce to the usual ones for $\gamma\approx1$). In particular, inflation may continue after $\phieps$, the inflaton eventually reaching (from a certain $\phiDBI$ onwards) an ultrarelativistic regime where $\gamma\gg1$. An analytical solution in the DBI dominated regime exists \cite{ourpaper}, which, however, is not inflating. This leaves the question if a significant amount of inflation is produced in the transitory regime $\phiDBI<\phi<\phieps$, which would affect the matching of today's scales to those during inflation. Access to this regime is through numerics only, and $\phieps$ and $\phiDBI$ are in fact of the same order, the number of e-folds produced inbetween typically being $\mathcal{O}(1)$. Hence, in the pure KKLMMT scenario, DBI dynamics do not significantly prolong inflation.

The second important point concerns the end of inflation: We do not forcibly have $\phiend=\phieps$, since inflation really ends at $\phistrg$, the onset of mutual brane annihilation. $\phistrg$ is calculated from the background parameters\footnote{Note that the dependence on $\alpha'$ cancels out.} $(\gs,M,v,\Nflux)$. Since $\phieps$ is known analytically, too, it is possible to express their ratio $\phieps/\phistrg=f(\gs,M,v,\Nflux)$, i.e.  as a function of background parameters.
$\phistrg$ could therefore lie ``on either side'' of $\phieps$, meaning that in some cases $\phiend=\phistrg$ while slow-roll still holds. Since $f(\gs,M,v,\Nflux)$ depends on the scale $M$, fixed from normalization to COBE (which, in turn, needs a $\phiend$ as an input), only the contour $\phieps/\phistrg = 1$ (at fixed $\gs$) can be traced unambiguously\footnote{COBE normalization is possible analytically when $\phiend=\phieps$, see \cite{ourpaper}.} in the parameter plane ($\ln v,\ln \Nflux$), see figure 1. Depending on the choice of $\gs$, some ($\ln v,\ln \Nflux$) belong to the region where $\phiend=\phistrg$, or where $\phiend=\phieps$. There exists, however, a rescaling of ($\Nflux,v,\gs$), illustrated by the lower panel in figure 1, that allows to remove the $\gs$ dependence. In the rescaled parameter space ($\ln x,\ln \bar{v}$), the contour $\phieps/\phistrg=1$ is unique.

We now turn to intrinsic parameter restrictions. First, consistency requires that the volume of the KS throat must not exceed the \emph{total} volume $V_{6}^{\mathrm{total}}$ of the 6d compactification \cite{volume}. Since $V_{6}^{\mathrm{total}}$ enters into the calculation of the 4d Planck mass, this constraint can be re-written as a condition relating $\mpl$ to ($\Nflux, v,\gs,\alphas$). This condition is a straight line with universal slope and $\alpha'$-dependent offset, cutting through the ($\ln x,\ln \bar{v}$) plane\footnote{See figure 6 of \cite{ourpaper}.}. Second, we focus on the case where inflation takes place in one throat: We require $\phiini<\phiuv$, and the throat has to be ``long enough'' to accommodate $\sim 60$ e-folds of inflation. In the region where $\phiend=\phieps$, this condition is another straight line, again with universal slope but an $\alpha'$-dependent offset. Where $\phiend=\phistrg$, the shape of this condition has to be found numerically.

\begin{figure}\label{parameters}
\includegraphics[width=4.5cm]{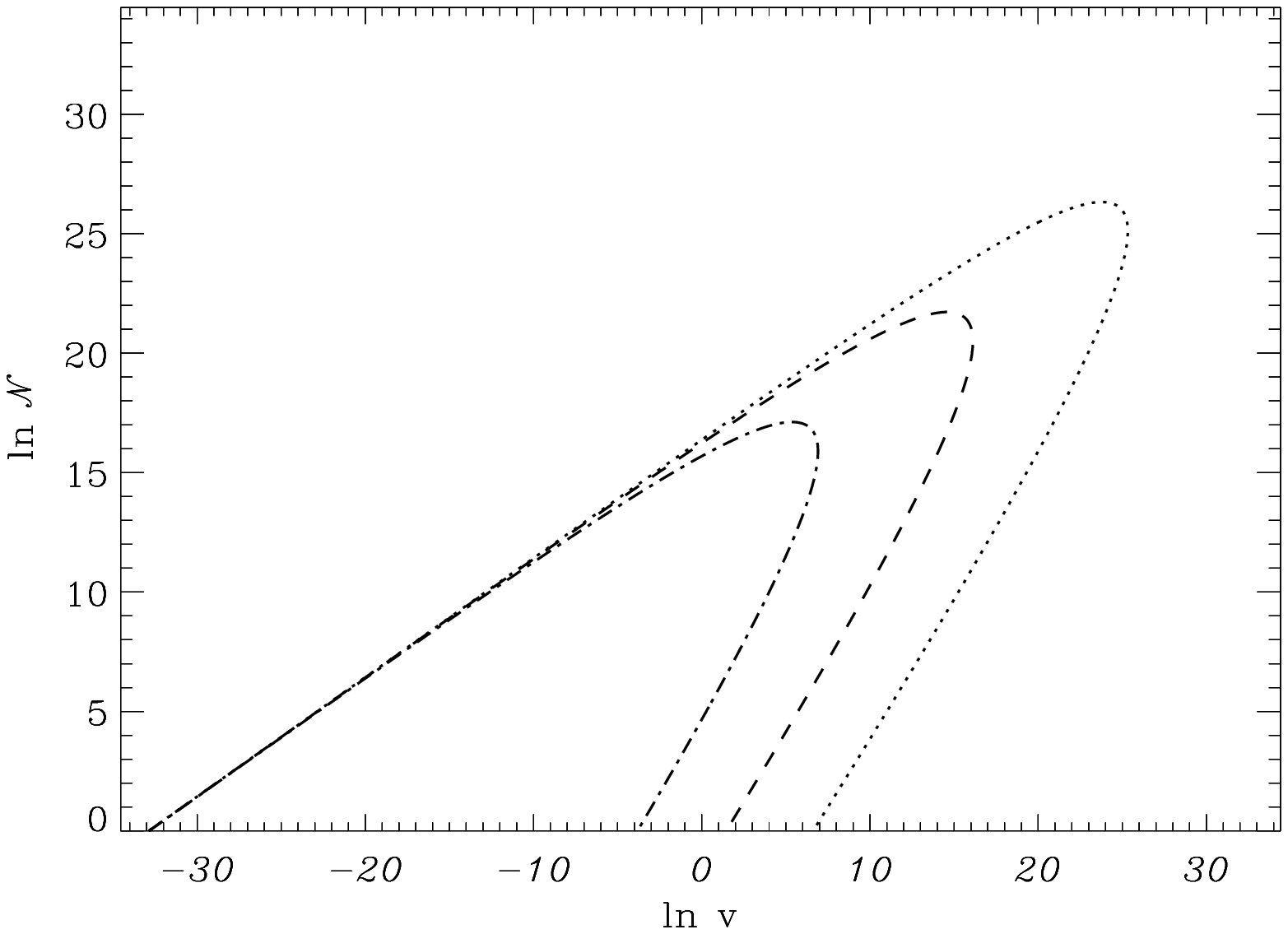}
\includegraphics[width=4.5cm]{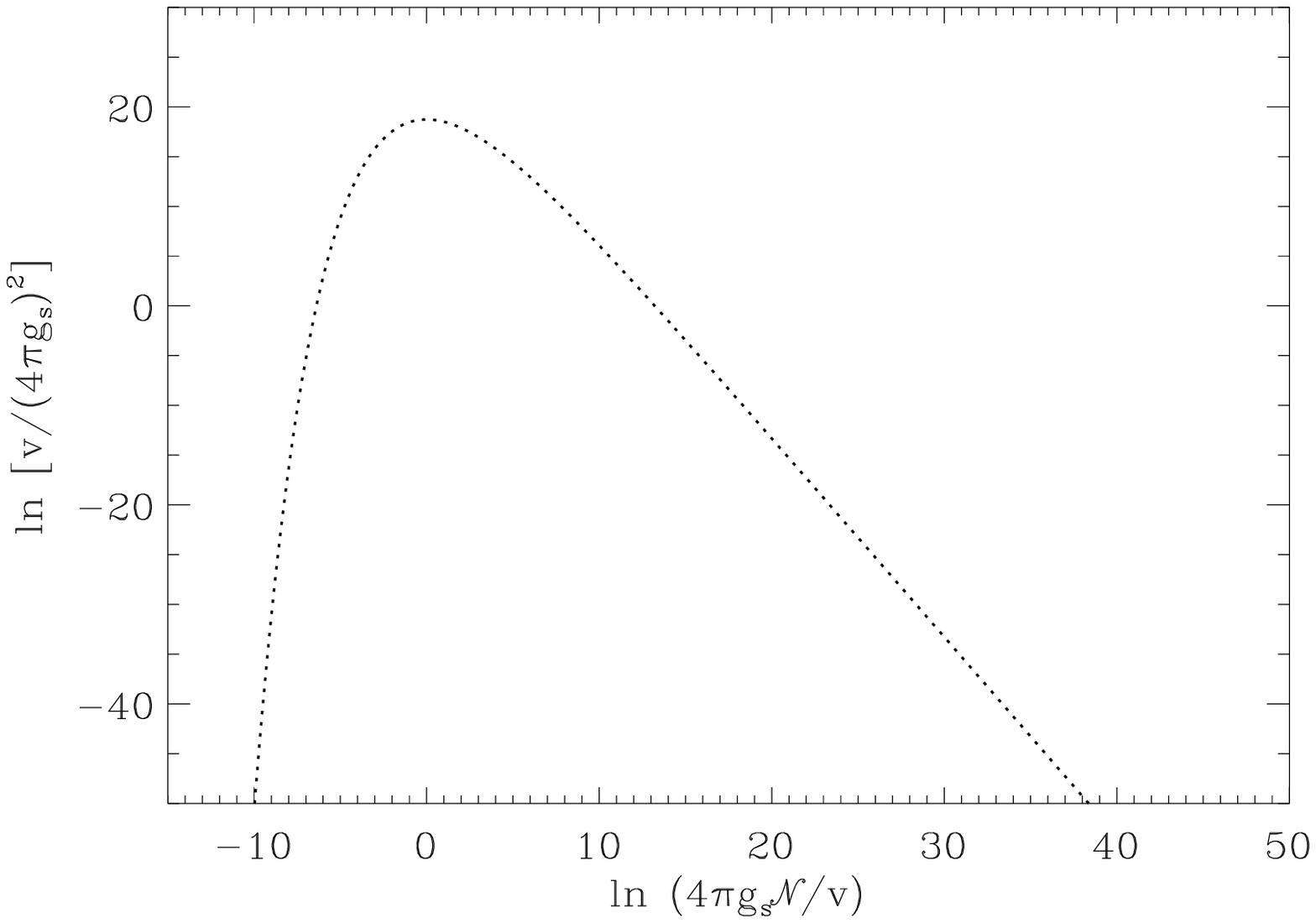}
\begin{minipage}[b]{6.8cm}
\caption{Upper panel: $\phieps=\phistrg$ in the plane ($\ln{\cal N}$, $\ln v$), using COBE normalization with $N_*=50$. The
  dotted line corresponds to $\gs=0.1$, dashed to
  $10^{-3}$ and dotted-dashed to $10^{-5}$. The area enclosed is the region where $\phieps>\phistrg$. The $\gs$-dependence can be absorbed by rescaling the
  parameters. Lower panel: $\phieps=\phistrg$ (universal for all values of $\gs$) in the plane ($\ln
  x,\,\ln\bar{v}$), where $x=4\pi\gs\Nflux/v$ and $\bar{v}=v/(4\pi\gs)^2$.}
\end{minipage}
\end{figure}

\section{MCMC results}
The KKLMMT model has four ``cosmological'' parameters $(M, \mu, \phistrg, \phiuv)$, to which we add the dimensionless parameter $R$ for the reheating era\footnote{The definition of (and prior on) $R$ is discussed in \cite{ourpaper}.}. The most suitable set for MCMC sampling, however, is $[\log\left(10^{10}\mathcal{P}_{*}\right),\,\log(\sqrt{\kappa}\mu),\,\log(\sqrt{\kappa}\phi_{_{\mathrm{UV}}}),\,\log(\phi_{\mathrm{strg}}/\mu),\,\ln R]$, where $\mathcal{P}_{*}$ is the amplitude of the scalar primordial spectrum at a given observable wavenumber $k_{*}$. Therefore, one has to implement the above restrictions as priors for these quantities. The numerics impose a lower limit of $\sqrt{\kappa}\mu>10^{-3}$. For a detailed discussion of all priors, see \cite{ourpaper}.

\begin{figure}
\includegraphics[width=7.5cm]{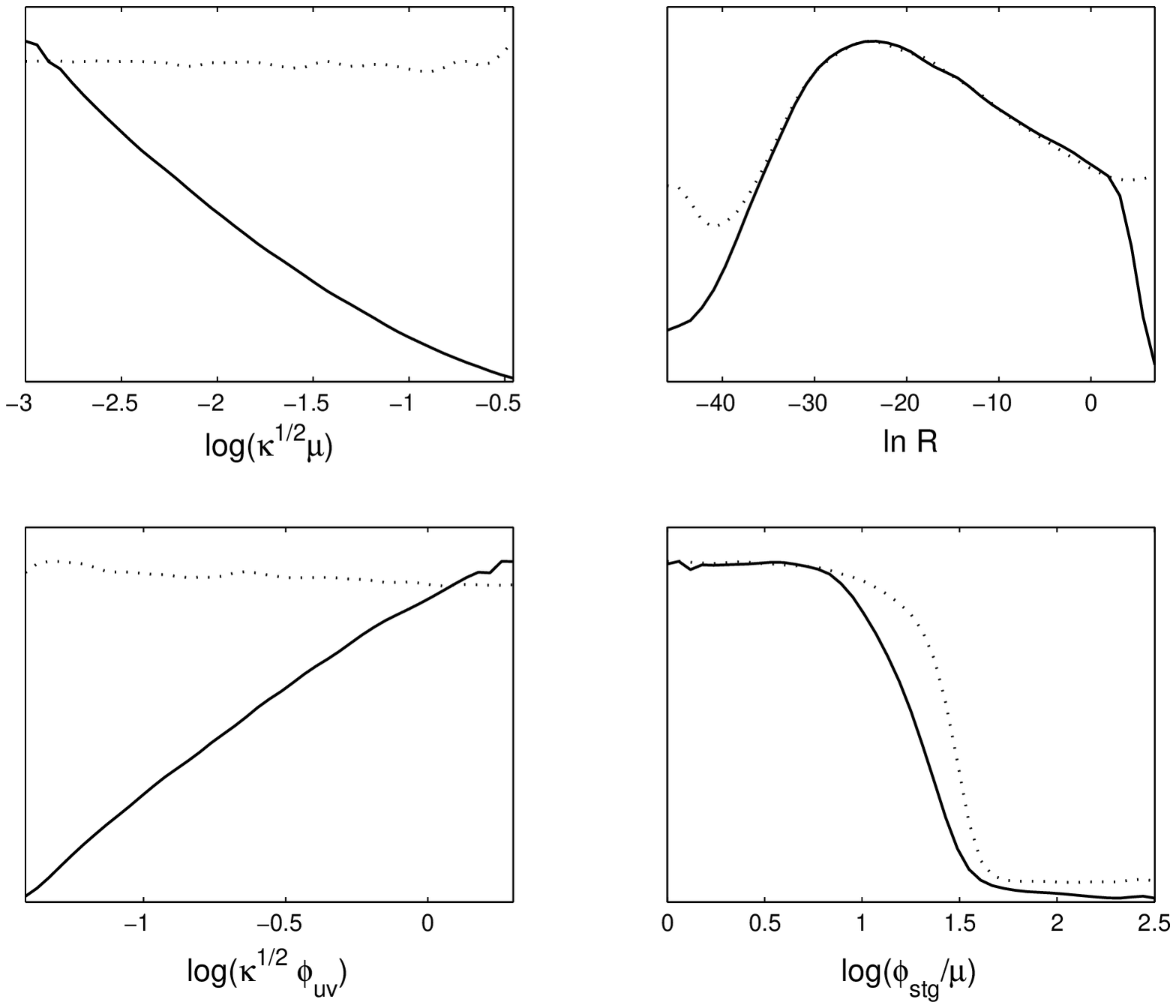}\hfill
\includegraphics[width=7.5cm]{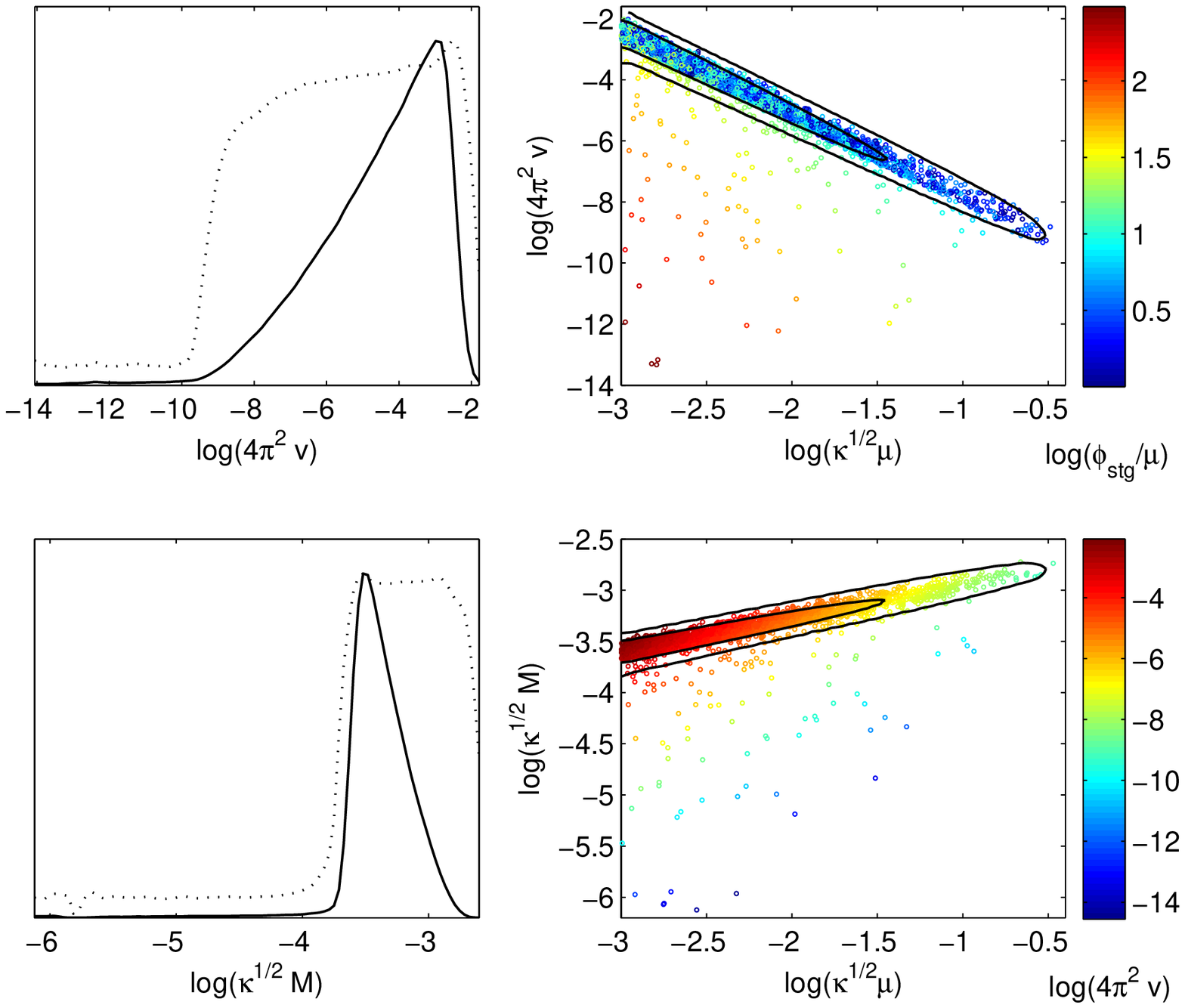}
\caption{\small{Left: MPD (solid
  lines) and ML (dotted lines) for the sampled
  primordial $\Lambda$CDM--KKLMMT parameters. Right: MPD and ML for $M/\mpl$ and $v$. On the very right are the $1\sigma$- and $2\sigma$-contours of the 2d posteriors obtained without marginalising
over $\log(\sqrt{\kappa}\mu)$. The 2d probability is
proportional to the point density while the colormap traces correlations
with the third parameter.}} 
\label{base} 
\end{figure}

We now briefly present the results of our MCMC comparison. First, one can show that the KKLMMT model reproduces $\Lambda$CDM parameters such as e.g. $\Omega_{\mathrm{b}},\Omega_{\mathrm{dm}},H_{0}$, as well as the correct perturbation amplitude and spectral indices. Second, figure \ref{base} shows the mean likelihoods (ML) and marginalized probability distributions (MPD) for the sampled primordial parameters $[\log(\sqrt{\kappa}\mu),\log(\sqrt{\kappa}\phi_{_{\mathrm{UV}}}),\log(\phi_{\mathrm{strg}}/\mu),\ln R]$. An interesting feature of the panels for  $\log(\sqrt{\kappa}\mu),\log(\sqrt{\kappa}\phi_{_{\mathrm{UV}}})$ is the \emph{difference} between ML and MPD: The ML's are uniform because in the explored prior range, these parameters do not improve the fit to the data, while the drop in the MPD's shows that $\log(\sqrt{\kappa}\mu)<1.1$ at 95\% confidence level (CL). These shapes are explained by volume effects in the multi-dimensional parameter space due to strong correlations. $\log(\phi_{\mathrm{strg}}/\mu)$ and $\ln R$, on the other hand, are directly constrained by the data: $\log(\phi_{\mathrm{strg}}/\mu)<1.4$ and $\ln R>-38$ at 95\% CL. Third, we can derive the corresponding distributions of the remaining parameters: $\log(\sqrt{\kappa}M)$ and $\mathcal{P}_{*}$ are directly related, as is $\log(4\pi^2 v)$ to $\mu$ and $M$ (see figure \ref{base}). In addition, the 2d probability distribution obtained without marginalising over $\log(\sqrt{\kappa}\mu)$ is shown, illustrating the strong correlations. In particular, the numerically motivated lower prior $\sqrt{\kappa}\mu>10^{-3}$ directly translates into an upper (lower) limit for $\log(4\pi^2 v)$ [$\log(\sqrt{\kappa}M)$]. The respective other end of these distributions, however, is ``physical'' and  gives the 95\% CL constraints  $\log(\sqrt{\kappa}M)<-2.9$ and $\log(4\pi^2 v)>-8.5$.

Do these probability distributions hold restrictions for fundamental string parameters, e.g.  $\gs$?  We know that $(M, \mu, \phistrg, \phiuv)$ really derive from the five quantities $(\gs,\alphas,M,v,\Nflux)$, hence additional assumptions are necessary. In \cite{ourpaper} this approach was explored, yielding a non-trivial degeneracy between $\Nflux,\gs$ and $v$ at a certain $\alphas$. However, the corresponding MPD would not allow any quantitative restriction on these parameters.

\section{Conclusions}
A general result of this work is that, \emph{in principle}, it seems possible to constrain stringy parameters from cosmology. However, the accuracy of present data does not suffice to break the degeneracies. Moreover, one must not underestimate the strong \emph{theoretical} prior that comes with any attempt at cosmological model building in string theory, since the testable inflationary quantities derive from fundamental (e.g. geometric) choices for the background.

In \cite{ourpaper}, we presented the first complete MCMC analysis of  the pure KKLMMT model, considering $\gs$ and $\alpha'$ as free parameters. We also suggest how to systematically scan the parameter space for arbitrary $\gs,\alpha'$. The data favour those cases where inflation occurs in the usual slow-roll way, i.e. inflation ends at $\phieps$ and not earlier at brane annihilation. This is because $\phiend=\phistrg>\phieps$ would push $n_{\mathrm{s}}\rightarrow 1$, while preserving a low level of gravitational waves (see \cite{ourpaper}). A weak limit on $v$, i.e. on a parameter of the 6d compactification, is also obtained: $\log v > -10$ at $95\%$ CL. 

The choice of the pure KKLMMT scenario \cite{kklmmt} comes with a considerable caveat: All moduli are considered stabilized, and various additional contributions to $V(\phi)$ are assumed to combine in such a way as to only leave the Coulomb term (\ref{potential}). Recently, these contributions became calculable \cite{corrections}, and their general cancellation is unlikely. In practice, the full potential should be of the form $V(\phi)=V_{D\bar{D}}(\phi)+m^{2}\phi^{2}+\dots$, leading to a completely different inflaton evolution and notably rendering the DBI phase important. The next step would be to include these
terms in our analysis, at the expense of introducing additional parameters.

\end{document}